\documentclass[twocolumn,aps,showpacs,showkeys,preprintnumbers,amsmath,amssymb]{revtex4}

\usepackage{graphicx}% Include figure files
\usepackage{dcolumn}% Align table columns on decimal point
\usepackage{bm}% bold math

\begin{document}

%\preprint{APS/123-QED}

\title{Tunable thermal emission at infrared frequencies via tungsten gratings}

\author{Jones T. K. Wan}\email{jwan@phy.cuhk.edu.hk}

\affiliation{Department of Physics, The Chinese University of Hong Kong,
Shatin, New Territories, Hong Kong.}

%\received{~~~~~~~~~~~~~~~~~~~~}

\date{\today}% It is always \today, today,
             %  but any date may be explicitly specified

\begin{abstract}
The author investigates the manipulation of thermal emission by using
one-dimensional tungsten gratings with different groove depths. It is
found that, by systematically increasing the depth of the groove, the
linearly polarized emission at particular frequencies can be  
substantially enhanced to achieve that of the blackbody radiation
limit, whereas the emission in other frequency ranges shows
no noticeable changes.  The results can provide useful insights into the
design of thermovoltaic applications.
\end{abstract}

% PACS, the Physics and Astronomy
%\pacs{Valid PACS appear here}% PACS, the Physics and Astronomy
                             % Classification Scheme.

%Use showkeys class option if keyword
%display desired
\maketitle

%%%%%% Introduction and review %%%%%%%%%%%%%%%%%%%%%%%%%%%%%%%%%%%%%%%%%%%%
Thermal radiation heat transfer is a ubiquitous process in nature, and
thermal emission has been an active research area for many
years \citep{planck_1959,howell_2002}. The main idea behind controlling the
emission spectrum of an object is to manipulate its photon density of
states (DOS). The photon DOS of a natural substance such as a metal or a dielectric,
can be modified slightly by changing the temperature. The
recent invention of man-made structures such as photonic crystals (PCs)
and plasmonic surfaces (PSs) has opened up the area of photon DOS control
\cite{joannopoulos_1995,sakoda_2001,raether_1988,chan_prl_2000,chan_opt_express_2001,PhysRevLett.86.4811,stefan.nat.photon.2007,billaudeau_apl_2008}.
Recent years have witnessed rapid growth in the study of thermal
radiation control via such artificial structures
\citep{hesketh_nature_1986,dowling_pra_1999,pendry_jpcm_1999,fleming_nature_2002,wan.apl.2006,han.prl.2007,lee.apl.2007,lee.opt.express.08,lee.adv.mater.2008,wang.opt.express.07,ikeda.apl.2008}.
For example, radiation at infrared frequencies can be
suppressed \citep{fleming_nature_2002} by manipulating the geometry and
material parameters of metallic PCs with large band gaps over the infrared
frequencies, thus allowing radiation at optical frequencies to be more
efficient. However, the complexity of the fabrication process for realistic
PCs remains a challenge. A structure that can be easily fabricated while
providing the flexible control of thermal radiation is therefore highly
desirable. Although the fabrication of bulk PC structures still presents a
major challenge, textured surfaces can be made relatively easily even for
sub-wavelength patterns. In addition, the advantages of bulk PCs over
surfaces with man-made structures in thermal radiation control are still
being debated. For instance, it is difficult to fabricate the woodpile
structure used by \citet{fleming_nature_2002} and \citet{han.prl.2007}
Although the proposed use of self-assembly
techniques \cite{velev_adv_mater_2000,JiangP._cm990080+} seems
promising, the tungsten inverse opal structure \cite{freymann.apl.2004}
fabricated by such a technique has not been shown to have substantial
superiority over other known PC structures in controlling thermal
emission. Although extensive attention is being focused on bulk PC structures,
the modification of thermal emission via textured surfaces has also been
widely studied.
%
%Recently, \citeauthor{fleming_nature_2002},\citep{fleming_nature_2002}
%\citeauthor{wan.apl.2006},\citep{wan.apl.2006} and
Recently, \citet{fleming_nature_2002}, \citet{wan.apl.2006}, and
\citet{yannopapas.prb.2006} reported independently that a thin photonic
crystal slab is sufficient to achieve the strong emission of electromagnetic
radiation at desired frequencies, and that emission at other frequencies
can be suppressed. Moreover, the thermal radiation effect of textured
surfaces was also studied by
%\citeauthor{pendry_jpcm_1999},\citep{pendry_jpcm_1999}
%\citeauthor{joulain.ssr.2005},\citep{joulain.ssr.2005} and
\citet{pendry_jpcm_1999}, \citet{joulain.ssr.2005}, and
\citet{fu.opt.lett.2005}. In view of these recent
findings, textured surfaces appear to be promising candidates for thermal
radiation control.
In this study, the author uses a model system consisting of tungsten
gratings and investigates the effects of groove depth on emission properties.
It is found that, by systematically increasing the height of the groove,
the emission at  particular frequencies can be tuned to achieve that
of the theoretical limit, that is, the blackbody radiation limit, wheres the
emission in other frequency ranges has no noticeable changes.
% Added on 2008.12.22
Therefore, the tungsten gratings may be useful to the design 
of thermal photovoltaic devices, which
require a structure that emits strongly at a particular frequency.
%

%%%%%% Technical details %%%%%%%%%%%%%%%%%%%%%%%%%%%%%%%%%%%%%%%%%%%%%%%%%%
%
%%%%%%%%%%% Figure 1: W grating
\begin{figure}[htb]
\includegraphics[width=0.9\columnwidth]{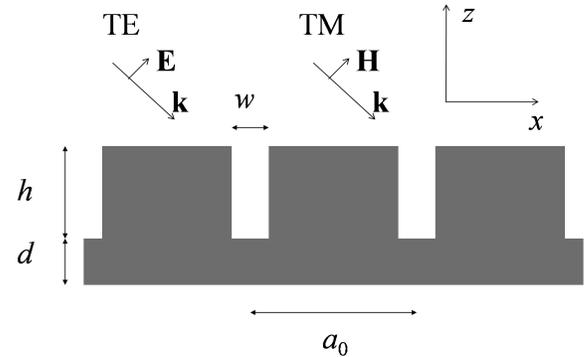}
\caption{Cross-session of the tungsten grating. The thickness ($d$)
is equal to 0.5 $\mu$m, and the other parameters are defined in the text.}
\label{grating}
\end{figure}
%%%%%%%%%%%%
%
The cross-section of the model tungsten grating is shown in
Fig.~\ref{grating}.
The grating has a period of $a_0 = 1.0$ $\mu$m and a thickness of $d = 0.5$
$\mu$m. It is considered to be optically thick because it is
much thicker than the typical skin depth ($\sim 20$ nm) of metals, thus allowing
no light to penetrate it.  The width of the groove is fixed at
0.3 $\mu$m, and the depths are fixed at $h = 0.5$ and 2.0 $\mu$m.
According to Kirchhoff's law, the spectral emittance $\epsilon_{\bf
k}(\omega)$ of a surface is equal to its spectral absorbance $A_{-{\bf
k}}(\omega)$ at thermal equilibrium; here, ${\bf k}$ and $\omega$ denote
the wavevector and angular frequency of the emitted radiation. The
validity of Kirchhoff's law for textured surfaces has been demonstrated by
\citet{joannopoulos_prl_2004}. Therefore, to understand the emission
properties of tungsten gratings,
the angle-dependent absorbances are calculated using
the transfer matrix method (TMM) \citep{bell_cpc_1995}.
The absorbance for each {\bf k} and $\omega$ is given by
$A_{-{\bf k}}(\omega) = 1- R_{\bf k}(\omega) - T_{\bf k}(\omega)$,
where $R_{\bf k}(\omega)$ and $T_{\bf k}(\omega)$ are the {\it total}
reflectance and transmittance of an incident wave.
In our calculations, $T_{\bf k}(\omega)=0$ because of the large thickness of
the gratings.

%%%%%% Results %%%%%%%%%%%%%%%%%%%%%%%%%%%%%%%%%%%%%%%%%%%%%%%%%%%%%%%%%%%%
%
%%%%%%%%%%% Figure 2: W dispersion diagram a1.0 w0.30.h0.50
\begin{figure}[htb]
\includegraphics[width=0.9\columnwidth]{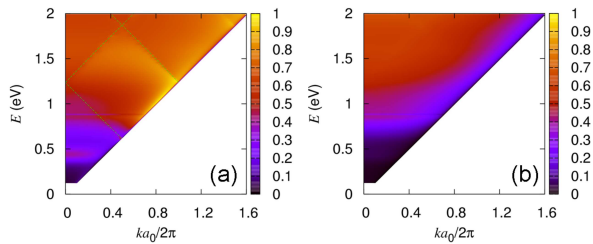}
\caption{
Calculated (a) TE and (b) TM angle-dependent absorbance spectra for a
grating of $h = 0.5$ $\mu$m. The high level of absorbance at $E > 0.9$ eV is mainly
due to the intrinsic absorption of bulk tungsten.  Surface plasmon
excitation [Eq. (\ref{dispersion.sp})] in TE polarization is indicated by
the green dashed lines. The dielectric data of tungsten is taken from
%\citeauthor{palik_1985} (see Ref. 30).
Ref. 31.
}
\label{band.1}
\end{figure}
%%%%%%%%%%%
%
%
%%%%%%%%%%% Figure 3: W dispersion diagram a1.0 w0.30.h2.00
\begin{figure}[htb]
\includegraphics[width=0.9\columnwidth]{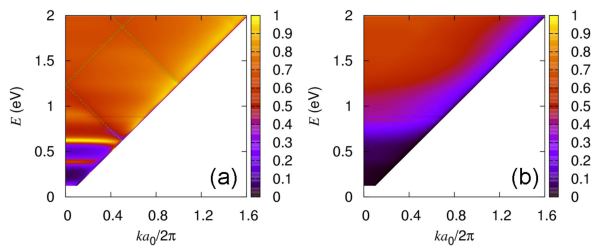}
\caption{
Calculated angle-dependent absorbance spectra for a grating of $h = 2.0$
$\mu$m. Note the absorption peaks at $E\sim$ 0.39 and 0.62 eV, which are due to the
excitation of the waveguide mode.
}
\label{band.2}
\end{figure}
%%%%%%%%%%%
%
The resultant absorbance spectra for TE
(${\bf E} \perp \hat{y}$) and TM (${\bf H} \perp \hat{y}$) polarizations
are shown in Figs.~\ref{band.1} and \ref{band.2}.
For the latter, no noticeable difference in absorbance between the different
structures can be observed, and the overall absorbance of both structures
is close to that of a plain tungsten slab.  The strong absorption for $E >
0.9$ eV ($\lambda \sim 1.38$ $\mu$m) is due to the intrinsic optical property
of tungsten \cite{rephaeli.apl.2008}. Therefore, the grating geometry does not
significantly enhance absorbance for TM polarization.  In
contrast, the absorbance for TE polarization changes considerably with
groove depth.  In addition to the strong tungsten absorption at $E > 0.9$ eV,
strong absorption peaks are also observed for TE polarization at $E > 1.2$ eV. The
locations of the absorption peaks appear consistently and systematically for
both $h = 0.5$ $\mu$m and $h = 2.0$ $\mu$m. In addition, the peaks exhibit typical
band-folding behavior, thus indicating that they are due to the excitation of
the propagating surface plasmon polaritons (SPPs) by the incident waves. To trace
the origin of the absorption, the dispersion relation of the propagating SPPs is
plotted on top of the absorbance spectra \citep{raether_1988}:
\begin{equation}
\frac{\omega}{c} =
\pm\sqrt{\frac{\epsilon_{\rm air}+\epsilon_{\rm W}}{\epsilon_{\rm air}\epsilon_{\rm W}}}
\bigg ( k + n\frac{2 \pi}{a_0}\bigg ),
\label{dispersion.sp}
\end{equation}
where $n$ is an integer, $\omega$ is the angular frequency, {\bf k} is the
in-plane wave vector in the $x$-direction, which is related to the incident
angle ($\theta$) by $k = (\omega/c)\sin\theta$,  $c$ is the speed of light, and
$\epsilon_{\rm air}$ and $\epsilon_{\rm W}$ are, respectively, the dielectric
functions of air and tungsten \citep{palik_1985}. As can be seen, equation
(\ref{dispersion.sp}) fits very well with the TE-absorption peaks for each
grating, which indicates that the physical origin of the absorption peaks for $E >
0.9$ eV is the excitation of the propagating SPPs.

We next focus on the region of $E < 0.9$ eV, in which absorption should be weak
for tungsten.  For $h= 0.5$ $\mu$m, non-dispersive absorption peaks are
observed at $\sim$ 0.45 eV ($\lambda\sim$ 2.8 $\mu$m).
As revealed later, the enhancement of absorption is
by almost a factor of 10 compared to that of a tungsten slab.  In addition, when the
depth increases from 0.5 $\mu$m to 2.0 $\mu$m, multiple peaks are observed at
$E = 0.39 $ eV ($\lambda\sim 3.2$ $\mu$m) and $E = 0.62 $ eV ($\lambda\sim 2.0$
$\mu$m).  Although not shown in this work, a similar scenario occurs if the groove
width increases, whereas no considerable absorption peaks are observed for
a groove width of less than 0.2 $\mu$m.  Moreover, it should be noted that at
$E\sim 0.62$ eV, the absorbance is very close to 1 and remains strong even when
the incident angle is close to 90$^\circ$.  These absorption peaks could be
%attributed to waveguide modes, as proposed by \citeauthor{PhysRevLett.83.2845},\citep{PhysRevLett.83.2845}
attributed to waveguide modes, as proposed by \citet{PhysRevLett.83.2845},
%which is due to the penetration of radiation into the groove.
in which a standing wave is formed inside the groove and results in energy
accumulation.

\begin{figure}[htb]
\includegraphics[width=0.9\columnwidth]{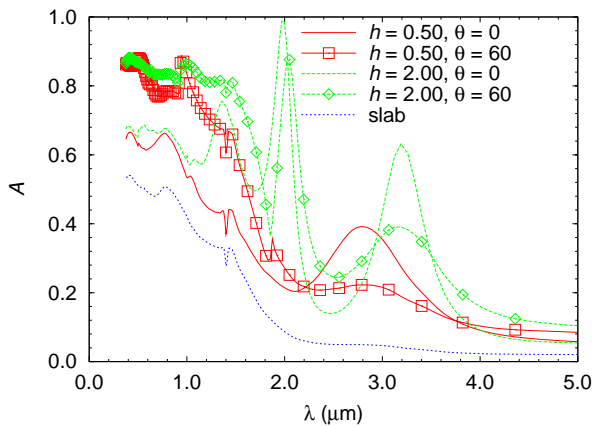}
\caption{
Absorbance spectra of tungsten gratings with different incident angles:
$\theta = 0^\circ$ (lines) and $\theta = 60^\circ$ (lines with symbols).
The normal absorbance spectrum of a 0.5 $\mu$m thick tungsten slab (blue
dashed line) is shown as a reference. The apparent minor irregularity in
the dispersion can be traced to corresponding irregularities in
experimental $\epsilon_{\rm W}$. The absorbance peaks at $\lambda \sim
2.8$ $\mu$m and $\lambda \sim 3.2$, 2.0 $\mu$m correspond to the peaks at
$E\sim 0.45$ eV in Fig.~\ref{band.1}a and $E\sim 0.39$, 0.62 eV in
Fig.~\ref{band.2}a.
}
\label{abs.spectra}
\end{figure}

To examine the dependence of TE absorbance on groove depth, the absorbance
spectra for each depth at two incident angles are plotted (0$^\circ$ and
60$^\circ$) in Fig.~\ref{abs.spectra}. The absorbance of a thick tungsten slab
is also drawn for comparison.  As can be seen, the absorbance of a flat tungsten slab
is low ($A < 0.02$) for near infrared (IR) wavelengths ($\lambda > 2 $
$\mu$m), which is due to the large dielectric mismatch between tungsten and
air.   As the magnitude of the real part of the dielectric function drops
significantly for $\lambda < 1.4$ $\mu$m, the dielectric mismatch is reduced,
thus resulting in the observed increase of absorbance for shorter wavelengths,
and $A$ reaches a plateau value of $A\sim 0.6$ for $\lambda < 0.8$ $\mu$m.
Therefore, the TM absorbances shown in Figs.~\ref{band.1} and \ref{band.2} could 
mainly be attributed to the intrinsic absorption of tungsten.

In the case of the TE mode, the effects of the groove can be clearly
seen. Depending on different geometries and incident angles, absorbance can
be enhanced by as much as a factor of 2 with respect to that of a tungsten slab.
For near IR wavelengths, the groove results in the great enhancement of
absorbance at certain frequencies.  For example, at normal incident ($\theta = 0^\circ$), $A$ is
enhanced by almost an order of 10 at $\lambda \sim 2.8$ $\mu$m for $h=0.5$ $\mu$m, which
corresponds to the peak at $E\sim$ 0.45 eV in Fig.~\ref{band.1}.  In
addition, a longer groove can result in more pronounced and sharper peaks;
this is evident from the absorbance of $h = 2.0$ $\mu$m. In such a case, $A$
has strong and non-dispersive peaks at $\lambda \sim 3.2$ and 2.0 $\mu$m,
respectively. The peaks at $\lambda\sim 2.0$ $\mu$m are very
close to 1 and are highly non-dispersive. The absorbance peak drops $\sim$
10\% only when the incident angle changes from 0$^\circ$ to 60$^\circ$, and
the peak position remains close to $\lambda\sim 2$ $\mu$m.

We now compare the TE thermal emission spectra of different gratings, and
the results are shown in Fig.~\ref{abs.spectra}.  The thermal emission spectra
are calculated by assuming Kirchhoff's law \cite{howell_2002}, and the thermal
emission spectrum $u_{\bf k}(\lambda,T)$ of wavevector ${\bf k}$ at
temperature $T$ is given by
\begin{equation}
u_{\bf k}(\lambda,T)=u_{b,{\bf k}}(\lambda,T)\times A_{\bf k}(\lambda),
\end{equation}
where $u_{b,{\bf k}}(\lambda,T)$ is the Planck spectrum of blackbody
radiation.
%
%%%%%%%%%%% Figure 4: W absorption spectrum
%%%%%%%%%%%
%
As our primary aim is to investigate the effect of the groove on the
radiation spectra, we ignore the temperature effect on the tungsten dielectric
constant \cite{arnold_ao_1984,saxena_pcm_1997}.

\begin{figure}[htb]
\includegraphics[width=0.9\columnwidth]{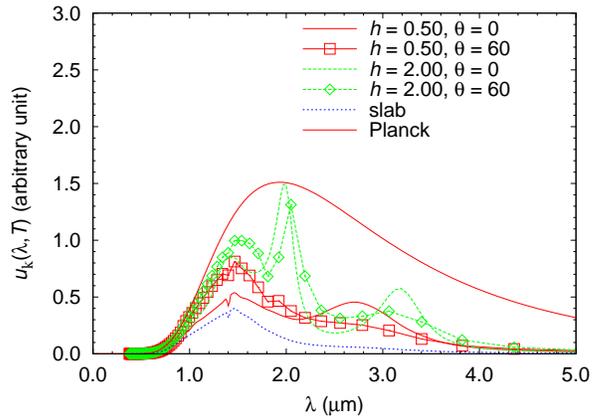}
\caption{
Thermal emission spectra of the tungsten gratings at 0$^\circ$ (lines) and
60$^\circ$ (lines with symbols)  at 1500 K. The corresponding flat slab
spectrum (blue dashed line) and Planck spectrum (red bold line) are also
shown.
}
\label{spectra.1500}
\end{figure}

\begin{figure}[htb]
\includegraphics[width=0.9\columnwidth]{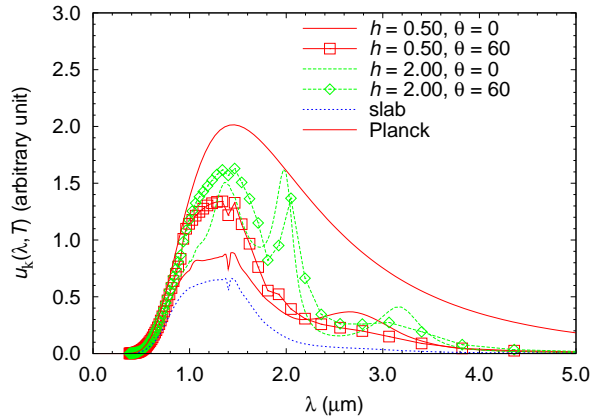}
\caption{
Thermal emission spectra at 2000 K.
}
\label{spectra.2000}
\end{figure}

In Fig.~\ref{spectra.1500}, the thermal emission spectra of the surfaces
corresponding to those in Fig.~\ref{abs.spectra} at 1500 K are drawn and
compared with that of a blackbody ($u_{\rm Planck}$).  In all cases, the
emission of the grating in all frequency ranges is enhanced with respect to that of
a slab, with the corresponding optical spectra being close to that of a blackbody.
Also, the peaks of the grating's emission spectra are effectively
red-shifted due to the strong absorption at infrared frequencies ($\lambda >
2$ $\mu$m), which is a direct consequence of the high level of absorbance
(Figs.~\ref{band.1}-\ref{band.2}).  Similar emission characteristics are
observed when the temperature is raised to 2000 K (Fig.~\ref{spectra.2000}).
The emission is now dominated by photons with shorter wavelengths (1.0 $\mu{\rm
m} < \lambda < 1.5$ $\mu${\rm m}), with the exception of $\lambda\sim 2.0$
$\mu$m for $h=2.0$ $\mu$m.  In other words, the thermal emission of photons at
$\lambda\sim 2.0$ $\mu$m is highly robust in all angles and a wide range of
temperatures ($\sim 500$ K).
%
%%%%%%%%%%% Figure 5: thermal emission spectra, T = 1500 K
%%%%%%%%%%%%
%
%
%%%%%%%%%%% Figure 6: thermal emission spectra, T = 2000 K
%%%%%%%%%%%%
%

The results shown in Figs.~\ref{band.1}-\ref{abs.spectra} suggest that
gratings are effective in manipulating the thermal emission of photons.
By controlling groove depth, the emission peak can be tuned to the
desired frequency. The reason for such emission enhancement could be
attributed to the excitation of the waveguide mode, in which thermally excited
photons at the resonant frequencies accumulate in the vicinity of
the groove. As the excited field is strong in the air, but weak on the
metallic surface, energy loss is minimal, and photons can be emitted.
% Added on 2008.12.22
To show the presence of the waveguide modes, we consider, for Fabry-P\'erot-like resonance,
the wavelengths of the $i$th and $j$th resonances are related by  
\begin{equation}
\frac{\lambda_i}{\lambda_j}\approx\frac{2j+1}{2i+1}.
\end{equation}
According to Figs. \ref{spectra.1500} and \ref{spectra.2000},
$\lambda \approx 3.2$ $\mu$m and 2.0 $\mu$m at resonance, which gives
$i=1, j=2, (2j+1)/(2i+1)=1.66$ and $\lambda_1/\lambda_2=1.6\approx 1.66$.
As the waveguide mode is highly frequency-selective, those thermally excited
photons that are not at the resonant frequencies are re-absorbed by
the groove and cannot be emitted.

It should be noted that the emission peak can be tuned according to the
operating temperature. For instance, the emission peak of blackbody
radiation is given by Wien's displacement law, $\lambda_{\rm max} =
b/T$, where $b = 2.897\times 10^{-3}$ m$\cdot$K is the Wien's displacement
constant. For $T=1500$ K, $\lambda_{\rm max} \sim 1.93$ $\mu$m, which is
close to the absorption peak of the $h=2.0$ $\mu$m grating shown in Fig.~\ref{band.2}.
Therefore, absorption spectra (Figs.~\ref{band.1}-\ref{abs.spectra})
provide useful information on how to achieve maximum emission at the
desired temperature.

%%%%%% Discussion and conclusion %%%%%%%%%%%%%%%%%%%%%%%%%%%%%%%%%%%%%%%%%%
In this study, the thermal property of emission is modified by enhancing the
emitting power of photons at frequencies at which the emitting power would be
low for a plain slab. Another way to manipulate thermal emission is to
suppress the emission of photons by introducing photonic or plasmonic band
gaps to the system.  The ultimate goal for thermal emission control is to
acheive both of these on the same geometry, that is, to enhance
emission in the desired frequency range while suppressing the emission of un-wanted
photons. Also, in practice, the intrinsic thermal properties
should be taken into account. For example, tungsten is used in this work
because of its high melting temperature ($\sim 3700$ K). Depending on the
operating temperature, one may choose an emitter made of other metals, and
the resultant emission properties will be different from those in this work.
Nevertheless, the principle of emission manipulation should remain the same.

The present study focuses on the effect of groove depth. However,
the period of the grating is also an important parameter for controlling
its optical properties. For example, the excitation of the SPPs
shown in Figs.~\ref{band.1}-\ref{band.2} is mainly controlled by
periodicity, and the excitation of SPPs via various geometries has been actively
studied.
% Added on 2008.12.22
In partcular, it has been shown that
\cite{greffet_nature_2002,yannopapas.prl.2007:053901}
the thermal radiation emitted via SPP should be highly directional and coherent,
and will resemble the state of coherence of laser radiation.
The tungsten gratings are thus expected to have 
SPP-originated emission properties similar to those
of the SiC grating of \citet{greffet_nature_2002}.
However, the absorbance spectrum shown in Fig. \ref{band.2}a is fairly broadband
even when SPPs are excited at $E > 0.6$ eV (see the green dashed lines in Fig. \ref{band.2}a),
and the emission is merely driven by the intrinsic absorption of bulk tungsten.
This should be due to the small periodicity used ($a_0 = 1.0$ $\mu$m) in this work
(compared to that of $\sim$ 6.25 $\mu$m in the work of \citet{greffet_nature_2002}).
One should be able to observe SPP-driven, directional emission when the periodicity is increased,
which will eventually lower the exicitation frequencies of SPP modes to frequencies below 0.6 eV. 
Nevertheless, the role of SPPs in the thermal emission of
grating will be the subject of further study.

Finally, the calculated absorbance spectra in this work can be verified
in the laboratory by measuring the zeroth order angle-dependence
reflectance. As the frequencies of interest are lower than the lowest SPP
excitation frequencies, the higher-order diffraction effect can be
neglected. However, the measurement of emission at high temperatures
remains a challenge. To conclude, our results suggest that thermal
emission can be controlled flexibly via metallic arrays of different
groove depths, thus resulting in an efficient infrared radiator. It is
hoped that this work will provide useful insights into the design of tunable
thermal emitting devices.

%%%%%% Acknowledgments and references %%%%%%%%%%%%%%%%%%%%%%%%%%%%%%%%%%%%
The author thanks Jensen Li and C. T. Chan for their discussions of this work
and acknowledges the support of S. S. Lam and T. L. Wan. Computation was
performed using the CUHK high-performance computing (HPC) facility. This
work is supported by RGC-HK (CERG project no. 403308).

%\bibliographystyle{epsl}
%\bibliographystyle{phaip}
%\bibliographystyle{apsrev}
%\bibliographystyle{plain}

%\bibliography{thermal}

\end{document}